\documentclass[12pt]{article}
\usepackage[colorlinks,allcolors=blue,bookmarks=false,hypertexnames=true]{hyperref}

\usepackage{fullpage,graphicx,psfrag,amsmath,amsfonts,verbatim,subcaption,url,booktabs}
\usepackage{mathtools}

\newcommand{\ones}{\mathbf 1}
\newcommand{\reals}{{\mbox{\bf R}}}

\newcommand{\Expect}{\mathop{\bf E{}}}
\newcommand{\Prob}{\mathop{\bf Prob}}

\newcommand{\eg}{{\it e.g.}}
\newcommand{\ie}{{\it i.e.}}

\newcommand{\BEAS}{\begin{eqnarray*}}
\newcommand{\EEAS}{\end{eqnarray*}}
\newcommand{\BEA}{\begin{eqnarray}}
\newcommand{\EEA}{\end{eqnarray}}
\newcommand{\BEQ}{\begin{equation}}
\newcommand{\EEQ}{\end{equation}}
\newcommand{\BIT}{\begin{itemize}}
\newcommand{\EIT}{\end{itemize}}

\title{Convex Optimization Over Risk-Neutral Probabilities}
\author{Shane Barratt \and Jonathan Tuck \and Stephen Boyd}

\begin{document}
\maketitle

\begin{abstract}
We consider a collection of derivatives that depend
on the price of an underlying asset at expiration or maturity.
The absence of arbitrage is equivalent to the existence of
a risk-neutral probability distribution on the price;
in particular, any risk neutral distribution can be interpreted 
as a certificate establishing that no arbitrage exists.
We are interested in the case when there are multiple
risk-neutral probabilities.
We describe a number of convex optimization problems over the
convex set of risk neutral price probabilities.
These include computation of bounds on the cumulative distribution,
VaR, CVaR, and other quantities, over the set of 
risk-neutral probabilities.
After discretizing the underlying price,
these problems become finite dimensional convex or quasiconvex 
optimization problems, and therefore are tractable.
We illustrate our approach using real options and futures pricing data for
the S\&P 500 index and Bitcoin.
\end{abstract}

\section{Introduction}
 
The arbitrage theorem
is a central result in finance
originally proposed by Ross 
\cite{ross1973return}.
For a market with a finite number of investments
and possible outcomes,
the arbitrage theorem states that there either exists a probability distribution
(called a \emph{risk-neutral probability})
over the outcomes such that the expected return
of all possible investments is nonpositive 
(\ie, arbitrage does not exist),
or there exists a nonnegative combination of the investments
that guarantees positive expected return 
(\ie, arbitrage exists).
The \emph{no-arbitrage assumption}
is that financial markets are arbitrage-free.
For the most part, this holds,
since if the markets were not arbitrage-free, 
someone would take advantage of the arbitrage,
changing the price until it no longer exists.
Under the no-arbitrage assumption,
a notable implication of the arbitrage theorem
is that a risk-neutral probability serves both as a 
conceivable distribution over the outcomes
and as a certificate ensuring 
that arbitrage is impossible.

For a given market,
the set of risk-neutral probabilities
is a polyhedron,
and arbitrage is impossible
if the set is nonempty.
We can verify that the market is arbitrage-free
by finding a point in the feasible
set of a particular system of linear equalities.
Apart from verifying the no-arbitrage assumption,
this set has many other uses.
For example, it has
been used for projecting onto the set of risk-neutral probabilities
using various distance measures
(\eg, $\ell_2$ norm, $\ell_1$ norm,
and KL-divergence) \cite{rubenstein1994estimation,jackwerth1996estimation,
stutzer1996estimation,buchen1996estimation,jackwerth1999option,branger2004pricing},
as well as for computing bounds on option prices given moments or
other information 
\cite{bertsimas1997introduction,bertsimas2002options,jackwerth2004option}.
These methods have been applied to various
derivative markets, including
equity indices \cite{bates2000post,aijo2008impact},
currencies \cite{castren2004options}, and
commodities \cite{melick1997recovering}.
We consider nonparametric models
of risk-neutral probabilities in this paper;
another viable option is to consider
parametric models, \ie, choose a distribution
and fit its parameters to observed pricing
data (see, \eg, \cite{bahra1a997implied,jackwerth1999option}
and the references therein).
We note that
once a risk-neutral distribution is found,
it is often used to construct
stochastic processes of
the price of the underlying asset,
\eg, as a binomial tree
\cite{rubenstein1994estimation,jackwerth1996generalized}.
Risk-neutral probabilities have also
been used to infer properties
of investor's utility functions \cite{ait2000nonparametric,jackwerth2000recovering}.

In this paper we consider the general problem of
minimizing a convex or quasiconvex function
over the (convex) set of risk-neutral probabilities.
By considering convex optimization problems,
finding a solution is tractable,
and indeed has linear complexity in the number
of outcomes,
which lets us scale the number of outcomes
to the tens of thousands.
Moreover, the advent of domain-specific languages (DSLs)
for convex optimization, \eg, CVXPY \cite{diamond2016cvxpy,agrawal2018cvxpy},
make not just solving, but also formulating these problems
straightforward; they require just a few lines
of a high-level language such as Python.
We show that there are many useful
applications of convex optimization
problems over risk-neutral probabilities,
which encompass a lot of prior work,
including computation of bounds
on expected values of arbitrary functions
of the expiration price,
estimation of the risk-neutral probability
using other information,
computation of bounds on the cost of existing
or new investments,
and sensitivities of various quantities
to the cost of each investment.
We illustrate a number of these applications
using real derivatives pricing data for
the S\&P 500 index and Bitcoin.

There are a number of notable limitations
to our approach.
First, we require the number of outcomes to
be finite and reasonably small.
Suppose, \eg, that we tried to apply our approach
to American-style options,
which can be exercised at any time up until to expiration.
Even if we discretized the price of the underlying asset
and time, the number of outcomes would be exponentially large,
since we would need to consider the price of the asset
at each time point until expiration.
(We note however that precise valuation and optimal exercise
of American options is still mostly an unsolved problem.)
Second, we consider static investments, \ie,
the investment is fixed until expiration.
This precludes multi-period investment models 
\cite{duffie2010dynamic},
dynamic hedging strategies that are at the core of 
derivative pricing models like the Black-Scholes 
model \cite{black1973pricing,merton1973theory},
as well as treatment of American options,
since we need to decide whether to exercise an option
or not based on the current price.
Despite these limitations, we find that our approach
can be very useful in practice and is also very interpretable,
as demonstrated by our examples in \S\ref{sec:examples}.

\paragraph{Outline.}
The remainder of the paper is organized as follows.
In \S\ref{sec:derivatives} we describe the setting
of the paper, define risk-neutral probabilities,
and give a characterization of the set of risk-neutral probabilities.
In \S\ref{sec:applications} we present the problem of convex
optimization over risk-neutral probabilities and give
a number of applications of this problem.
Finally, in \S\ref{sec:examples}, we illustrate our approach
on real derivatives pricing data for the S\&P 500 Index and Bitcoin.

\section{Risk-neutral probabilities}\label{sec:derivatives}

\paragraph{Setting.}
We consider a market for an asset, referred to as the \emph{underlying},
with a number of derivatives 
that provide payoffs at the same future date or time,
referred to as expiration or maturity.
We assume that there are $n$ possible investments that
include, \eg, buying or (short) selling
the underlying, as well as buying
or selling (writing) derivatives.
We let $p>0$ denote the price
of the underlying at expiration.

\paragraph{Payoff.}
The payoff function $f_i:\reals_+\to\reals$ denotes the dollar amount
received (or paid, if negative) per unit held
of the $i$th investment;
we give some examples of payoff functions below.
If we own a quantity $w_i\geq 0$ of the
$i$th investment, then at expiration, we would receive
$f_i(p)w_i$ dollars.
(We note that we do not discount payoffs at the risk-free rate,
but we could easily do this in our formulation.)

\paragraph{Cost.}
We let $c\in\reals^n$
denote the cost in dollars
to acquire one unit of each investment
($c_i<0$ means that we are paid to acquire the investment).
The cost is the ask price if we are purchasing
and the negative bid price if we 
are selling, adjusted for fees and rebates.
If we acquired a quantity $w_i\geq 0$ of the
$i$th investment, it would cost us $c_iw_i$ dollars,
and the return of our investment, at expiration,
would be $(f_i(p) - c_i) w_i$ dollars.

\subsection{Examples of payoff functions}

In this section we give some examples of payoff functions
(see, \eg, \cite{GVK563580607} for an overview of various derivatives).

\paragraph{Underlying.}
In many cases we can directly invest in the underlying.
We allow both going long (buying), and going short (selling borrowed shares).
Going long in the underlying has
a payoff function 
\[
f(p) = p,
\]
and going short in the underlying has the payoff function
\[
f(p) = -p.
\]

\paragraph{European options.}
A European option is a contract that gives one party the right
to buy or sell an underlying asset at an agreed upon strike price.
If the right is to buy the underlying asset (a call option), the option will only
be exercised if the underlying price is greater than the strike price.
Conversely, if the right is to sell the underlying asset (a put option),
the option will only be exercised if the underlying price is less
than the strike price.
Under this logic, the payoff functions for buying European options with
a strike price $s$ are
\[
f^\mathrm{call}(p) = (p-s)_+, \quad f^\mathrm{put}(p) = (p-s)_-,
\]
where $x_+ = \max(x, 0)$ and $x_- = (-x)_+$.
The payoff function for selling (writing) a European option is
\[
f^\mathrm{w,call}(p) = -f^\mathrm{call}(p), \quad f^\mathrm{w,put}(p) = -f^\mathrm{put}(p).
\]

\paragraph{Futures.}
Futures are contracts that obligate the buyer of
the contract to buy or sell 
the underlying asset at an agreed upon strike price.
A long futures contract means the party must buy
the underlying asset at that strike price.
Denoting the strike price of the future by $s$, the payoff
for buying a long futures contract is
\[
f(p) = p - s.
\]
A short futures contract means the party must sell
the underlying asset at that strike price.
The return function for buying a short futures contract is
\[
f(p) = s - p.
\]

\paragraph{Binary options.}
A binary option is a contract that pays either a fixed
monetary amount or nothing depending on
the underlying's price.
For example, a binary option that pays the buyer
one dollar if the underlying
asset is above a strike price $s$
has a payoff function
\[
    f(p) = 
    \begin{cases}
        1 & p \geq s,\\
        0 & \text{otherwise}.
    \end{cases}
\]
If we sell that same binary option, the payoff function is
\[
    f(p) = 
    \begin{cases}
        -1 & p \geq s,\\
        0 & \text{otherwise}.
    \end{cases}
\]

\subsection{Discretized outcomes}
For the remainder of the paper we will work with a discretized version
of the price $p$,
meaning it can only take one of $m$ values $p_1,\ldots,p_m$,
where we assume $p_1 < p_2 < \cdots < p_m$.
(We note that the discretization can be unequally spaced.)
Since the methods that we describe involve
convex optimization, they scale well (and often linearly) with $m$~\cite{BoV:04};
this implies that $m$ can be chosen to be large enough that the 
discretization error is negligible.

We can define a probability distribution
over $p$ as a vector $\pi \in \reals^m$, with
$\Prob(p = p_i) = \pi_i$.
Such a vector is in the set
\[
\Delta=\{\pi \in \reals^m \mid \pi \geq 0,~ \ones^T\pi=1\},
\]
\ie, the probability simplex in $\reals^m$.

\subsection{The set of risk-neutral probabilities}

\paragraph{Payoff matrix.}
We can summarize the payoffs of each investment
for each possible outcome with the payoff matrix $P\in\reals^{m\times n}$,
with entries given by
\[
P_{ij} = f_j(p_i), \quad i=1,\ldots,m, \quad j=1,\ldots,n.
\]
Here $P_{ij}$ is the payoff in dollars per unit invested
in investment $j$, if outcome $i$ occurs.

\paragraph{Arbitrage.}
Let $w\in\reals_+^n$ denote an investment vector,
meaning we invest in a quantity $w_i$
of the $i$th investment,  and hold these
investments until expiration.
The overall investment will cost us $c^Tw$
now,
and our expected payoff at expiration will be $\pi^TPw$,
meaning our expected return
is $(P^T\pi-c)^T w$.
Arbitrage is said to exist if there exists an investment
vector that guarantees positive expected return,
\ie, there exists $w \geq 0$ with $(P^T\pi - c)^T w>0$.
Equivalently, arbitrage is said to exist if the homogeneous linear program (LP) 
\BEQ
\begin{array}{ll}
\mbox{maximize} & (P^T\pi - c)^T w\\
\mbox{subject to} 
& w \geq 0,
\end{array}
\label{eq:arbitrage_primal}
\EEQ
with variable $w$, is unbounded above.

\paragraph{The set of risk-neutral probabilities.}
We say that $\pi$ is a \emph{risk-neutral probability}
(or \emph{no-arbitrage distribution})
if arbitrage is impossible,
that is, if the optimal value of problem~\eqref{eq:arbitrage_primal} is bounded.
By LP
duality \cite{dantzig1998linear} or the Farkas lemma \cite{farkas1902theorie},
\eqref{eq:arbitrage_primal} is bounded
if and only if $P^T\pi \leq c$.
This means that the \emph{set of risk-neutral probabilities} is
the (convex) polyhedron
\[
\Pi = \{\pi \in \Delta \mid P^T\pi \leq c\}.
\]
We note that if $\Pi$ is empty, then arbitrage exists.
We can interpret $\pi\in\Pi$ as a distribution
over the outcomes for which it is impossible to
invest and receive positive expected return.

\paragraph{Another interpretation.}
Consider the problem
\BEQ
\begin{array}{ll}
\mbox{maximize} & t\\
\mbox{subject to} 
& Pw - (c^Tw)\ones \geq t\ones,\\
& w \geq 0,
\end{array}
\EEQ
with variable $w$.
(This problem is equivalent to problem~\eqref{eq:arbitrage_primal}.)
If it is unbounded above, then for every $R>0$,
there exists an investment vector $w$ that guarantees
our return will be at least $R$, no matter what the outcome is.
The dual is
\BEQ
\begin{array}{ll}
\mbox{maximize} & 0\\
\mbox{subject to} 
& \pi \in \Pi,
\end{array}
\EEQ
with variable $\pi$.
Therefore, another interpretation of $\pi\in\Pi$ is as a \emph{certificate}
guaranteeing that it is impossible to always have positive return
regardless of the outcome.

\begin{figure}
\centering
\includegraphics[width=.4\textwidth]{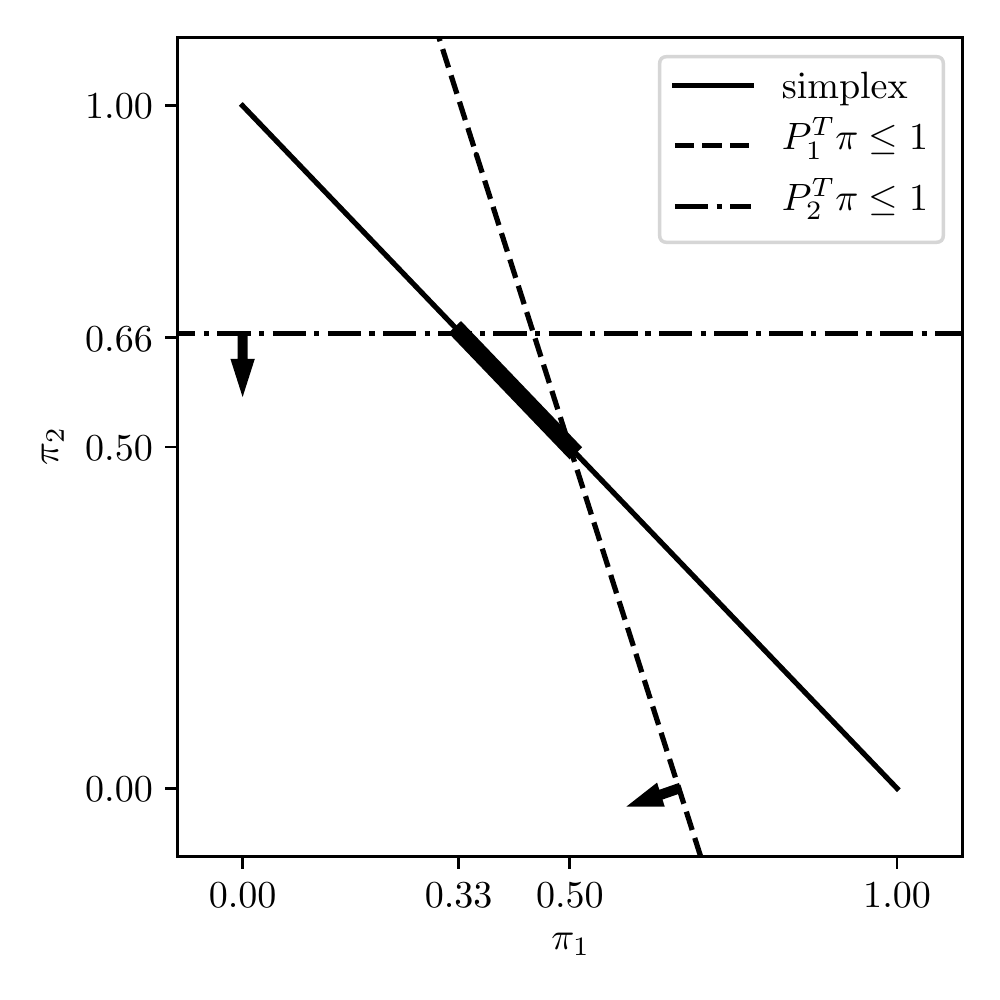}
\caption{An example of the set of risk-neutral probabilities $\Pi$, 
denoted by the thick line segment. 
Here $P_i$ denotes the $i$th column of $P$.}
\label{fig:example}
\end{figure}

\paragraph{Example.}
Suppose there are $n=2$ investments, $m=2$ outcomes,
the prices are $c=(1,1)$, and
the payoff matrix is
\[
P = \begin{bmatrix}
3/2 & 0 \\
1/2 & 3/2
\end{bmatrix}.
\]
Then the set of risk-neutral probability distributions is
\[
\Pi = \{(x, 1-x) \mid 1/3 \leq x \leq 1/2\}.
\]
We visualize the construction of this set in figure~\ref{fig:example}.

\section{Convex optimization over risk-neutral probabilities}\label{sec:applications}

The general problem of convex optimization over risk-neutral probabilities
is
\BEQ
\begin{array}{ll}
\mbox{minimize} & L(\pi) \\
\mbox{subject to} 
& \pi \in \Pi,
\end{array}
\label{eq:prob}
\EEQ
with variable $\pi$,
where $L:\reals^m\to\reals\cup\{+\infty\}$ is convex
(or quasiconvex).
We use infinite values of $L$ to encode constraints.
$\Pi$ is a polyhedron, so~(\ref{eq:prob}) is a convex optimization problem \cite{BoV:04}.
In general problem~(\ref{eq:prob}) does not have
an analytical solution,
but we can numerically find the global optimum efficiently
using modern convex optimization solvers \cite{BoV:04}.
All of the problems we describe below (and many others) 
are readily expressed in a few lines
of code using domain specific languages for convex optimization,
such as CVX \cite{grant2008cvx,grant2014cvx}, 
CVXPY \cite{diamond2016cvxpy,agrawal2018cvxpy}, 
Convex.jl \cite{udell2014convexjl}, 
or CVXR \cite{fu2019cvxr}.

\subsection{Functions of the price}
\label{sec:funcs}
Suppose $g:\reals\to\reals$ is some function
of the underlying's price at expiration;
the expectation of $g$ is
\[
\Expect g(p) = \sum_{i=1}^m \pi_i g(p_i),
\]
which is a linear function of $\pi$.
Some examples of functions of the price include:
\begin{itemize}
    \item \emph{The price.} Here $g(p)=p$. The expected value is the expected price.
    \item \emph{The return on an investment.} Here $g(p) = \sum_{i=1}^n (f_i(p) - c_i)w_i$ for
    an investment $w\in\reals_+^n$. The expected value is the expected return of
    the investment.
    \item \emph{Indicator functions of arbitrary sets.} Here $g(p) = 1$ if $p\in C$ and 0 otherwise,
    for some set $C\subseteq\reals$. The expected value is $\Prob(p \in C)$.
\end{itemize}

\paragraph{Bounds on expected values.}
We can compute lower and upper bounds on
expected values of functions of the price
by respectively letting $L(\pi)=\Expect g(p)$
and $L(\pi) = -\Expect g(p)$
and solving problem \eqref{eq:prob}.
For example, we could compute bounds on
the expected price or the return
on a given investment.

\paragraph{Bounds on ratios of expected values.}
If we have another function $f$
of the price, and $g(p) > 0$,
then the function
\[
\frac{\Expect f(p)}{\Expect g(p)} = \frac{\sum_i \pi_i f(p_i)}{\sum_i \pi_i g(p_i)},
\]
is quasilinear.
We can find bounds on this ratio by minimizing and maximizing this
quantity, both of which are both quasiconvex optimization problems.
For example, we can compute bounds on
$\Prob(p \in A \mid p \in B)$ for two sets $A\subseteq\reals$ and $B\subseteq\reals$,
since it is equal to
\[
\frac{\Prob(p \in A \cap B)}{\Prob(p \in B)}.
\]

\paragraph{CDF.}
The cumulative distribution function (CDF) of $g$ is the function
\[
F(x) = \Prob(g(p) \leq x) = \sum_{g(p_i) \leq x} \pi_i,
\]
which, for each $x$, is linear in $\pi$.
For example, if $g(p)=p$, then $F(x)$
is just the CDF of the price.
We can compute lower and upper bounds on the CDF
at $x=p_1,\ldots,p_m,$
by minimizing and maximizing $F(x)$ subject to $\pi\in\Pi$.

\paragraph{VaR.}
The value-at-risk of $g(p)$ at probability $\epsilon\in[0,1]$ is
defined as
\[
\mathbf{VaR}(g(p);\epsilon) 
= \inf\{\alpha \mid \Prob(g(p) \leq \alpha) \geq \epsilon\} = F^{-1}(\epsilon),
\]
where $F^{-1}(\epsilon)=\inf\{x \mid F(x) \geq \epsilon\}$~\cite{duffie1997VaR}.
From the bounds on the CDF, we can compute bounds on the value
at risk as
\[
F_\mathrm{max}^{-1}(\epsilon) \leq \mathbf{VaR}(g(p);\epsilon) \leq F_\mathrm{min}^{-1}(\epsilon).
\]

\paragraph{CVaR.}
The conditional value-at-risk of $g(p)$ at probability
$\epsilon$ is defined as (see, \eg, \cite{rockafellar2000optimization})
\[
\mathbf{CVaR}(g(p);\epsilon) 
= \inf_{\beta} \left(\beta + \frac{\Expect(g(p)-\beta)_+}{1-\epsilon}\right)=
\min_i \left(p_i + \sum_{j=1}^m \pi_j \frac{(g(p_j)-p_i)_+}{1-\epsilon}\right),
\]
which is a concave function of $\pi$.
Therefore, we can find an upper bound on $\mathbf{CVaR}$
by letting $L(\pi) = -\mathbf{CVaR}(g(p);\epsilon)$.
Since conditional value-at-risk is bounded below by value-at-risk,
$F_\mathrm{max}^{-1}(\epsilon)$ is a (trivial) lower bound.

\paragraph{Constraints.}
We can incorporate upper or lower bounds on
the expected values of functions of the price as linear equality
constraints in the function $L$.
These linear inequality constraints can be interpreted as adding
another investment.
For example, if we add the constraint $a^T\pi \leq b$ for $a\in\reals^m$ and 
$b\in\reals$, this is the same as if we had originally included an
investment with a payoff function $f(p_i)=a_i$ and cost $b$.

\subsection{Estimation}

\paragraph{Maximum entropy.}
We can find the maximum entropy risk-neutral probability
by letting
\[
L(\pi)=\sum_{i=1}^m \pi_i \log(\pi_i).
\]

\paragraph{Minimum KL-divergence.}
Given another distribution $\eta \in \Delta$,
we can find the closest risk-neutral probability
distribution to $\eta$ as measured
by Kullback-Leibler (KL) divergence by letting
\[
L(\pi) = \sum_{i=1}^m \pi_i \log(\pi_i/\eta_i).
\]

\paragraph{Closest log-normal distribution.}
We can approximately find the closest log-normal distribution
to $\Pi$ by performing the following alternating
projection procedure, starting with $\pi_0\in\Pi$:
\begin{itemize}
    \item Fit a log-normal distribution to $\pi_k$ with mean and variance
    \[
    \mu = \sum_{i=1}^m \pi_i \log(p_i), \quad \sigma^2 = \sum_{i=1}^m \pi_i (\log(p_i) - \mu)^2.
    \]
    \item Discretize this distribution, resulting in $\eta_k\in\Delta$.
    \item Set $\pi_{k+1}$ equal to the closest risk-neutral probability distribution 
    to $\eta_k$, in terms of KL-divergence. If $\pi_{k+1}$ is close enough to $\eta_k$, then quit.
\end{itemize}
For better performance, this process may be repeated for various $\pi_0 \in \Pi$.

\subsection{Bounds on costs}

Suppose that we want to add another investment,
and would like to come up with
lower and upper bounds on its cost
subject to the constraint that
arbitrage is impossible, \ie,
there exists a risk-neutral probability distribution.
Suppose the
payoff function of the new investment
is $f(p_i) = (p_\mathrm{new})_i$,
where $p_\mathrm{new}\in\reals^m$.
We can find lower and upper bounds
on the cost of this new investment
by respectively letting
$L(\pi) = p_\mathrm{new}^T\pi$
and $L(\pi) = -p_\mathrm{new}^T\pi$
and solving problem \eqref{eq:prob}.
Bertsimas and Popescu \cite[\S3]{bertsimas2002options}
were among the first to propose computing bounds on option prices
based on prices of other options.

\paragraph{Validation.}
We can check whether our prediction is accurate
by holding out each investment
one at a time and comparing
the lower and upper bounds that we find with the
true price.

\subsection{Sensitivities}
Suppose $L$ is convex and
let $\lambda^\star\in\reals_+^n$ denote the optimal dual variable
for the constraint $P^T\pi \leq c$ in problem \eqref{eq:prob},
and let $L^\star(c)$
denote the optimal value as a function of $c$.

\paragraph{A global inequality.}
For $\Delta c\in\reals^n$, the following global inequality holds \cite[\S5.6]{BoV:04}:
\[
L^\star(c + \Delta c) \geq L^\star(c) - (\lambda^\star)^T \Delta c.
\]

\paragraph{Local sensitivity.}
Suppose $L^\star(c)$ is differentiable at $c$.
Then $\nabla L^\star(c) = -\lambda^\star$ \cite[\S5.6]{BoV:04}.
This means that changing the costs of the investments
by $\Delta c\in\reals^n$ will decrease $L^\star$ by roughly $(\lambda^\star)^T\Delta c$.

\section{Numerical examples}\label{sec:examples}

We implemented all examples using CVXPY
\cite{diamond2016cvxpy,agrawal2018cvxpy},
and each required just a few lines of code.
The code and data for all of these examples
have been made freely available online at
\[
\verb|www.github.com/cvxgrp/cvx_opt_risk_neutral|
\] 

\subsection{Standard \& Poor's 500 Index}
In our first example, we consider the
Standard \& Poor's 500 index (SPX) as the underlying,
which is a market-capitalization-weighted index
of 500 of the largest publicly traded U.S. companies,
and excludes dividends.
We gathered the end-of-day (EOD)
best bid and ask price of all SPX
options on June 3, 2019,
as well as the price of the index,
which was 2744.45 dollars,
from the OptionMetrics Ivy database via the
Wharton Research Data Services~\cite{WRDS}.

We discretized the expiration price
from 1500 to 4000 dollars, in 50 cent increments,
resulting in $m=5000$ outcomes.
We allowed six possible investments:
buying or selling puts, 
buying or selling calls, and
buying or selling the underlying.
The payoffs for each of these investments are 
described in \S\ref{sec:derivatives}.
The cost of each investment is
the ask price if buying, the negative bid price if selling,
plus a 65 cent fee for buying/selling each option
(which at the time of writing are the fees
for the TD Ameritrade brokerage),
and a $0.3\%$ fee for buying or selling the underlying.

We consider the options that
expire 25 days into the future, on June 28, 2019.
There were 112 puts and 81 calls expiring on June 28
that had non-empty order books, \ie,
had at least one bid and ask quote.
Therefore, we allow $n=2(112+81)+2=388$ investments.

\paragraph{Functions of the price.}
We calculated bounds on the expected value of the expiration
price. The lower bound was 2745.77 dollars and the upper bound 
was 2747.03 dollars.
We then computed bounds on the probability that the expiration price
is $20\%$ below the current price, given that the expiration price
is less than the current price; this probability was
found to be between $0.4\%$ and $2\%$.
We also computed bounds on the CDF, complementary CDF (CCDF),
and VaR of the expiration price,
and plot these figure~\ref{fig:bitcoin_cdf_var} we plot these bounds.

\begin{figure}
\centering
\begin{subfigure}{0.8\textwidth}
  \includegraphics[width=\linewidth]{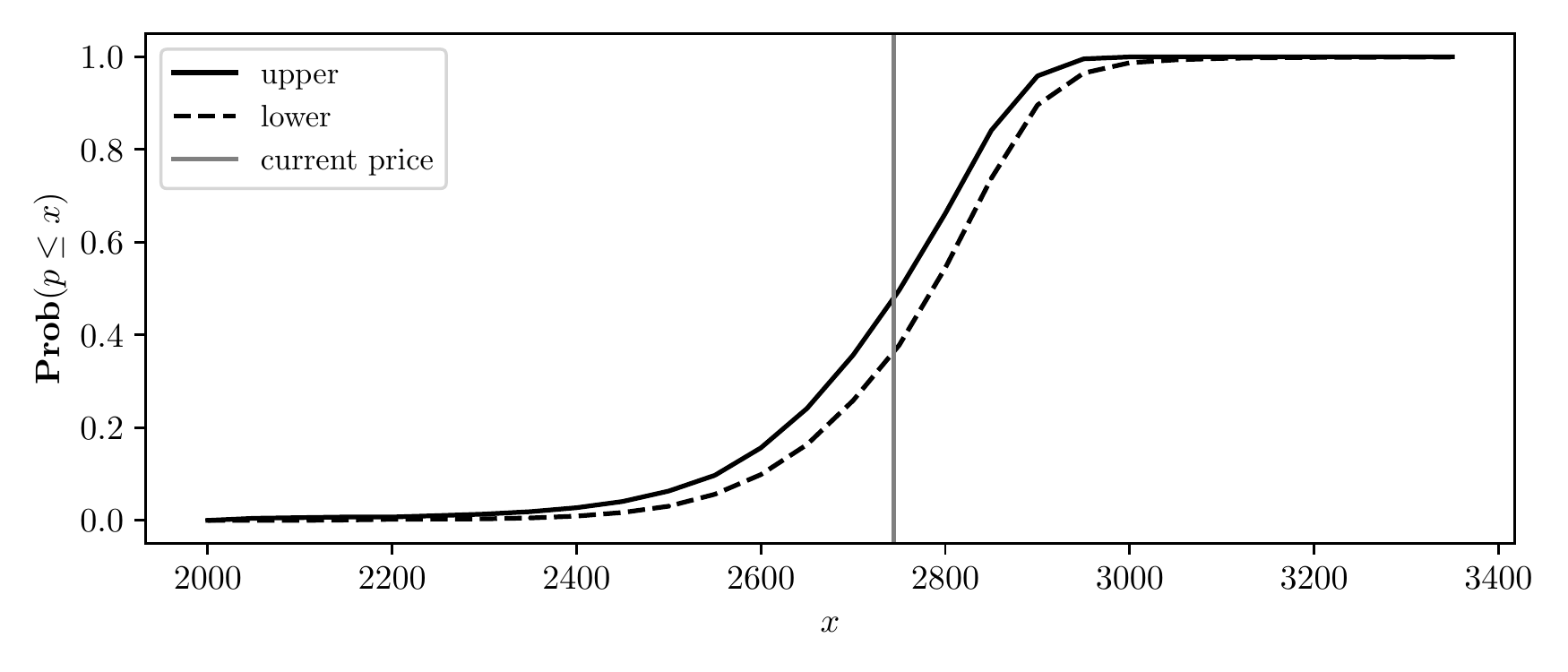}
  \caption{Bounds on $\Prob(p \leq x)$.}
  \label{fig:3}
\end{subfigure}\\

\medskip
\begin{subfigure}{0.8\textwidth}
  \includegraphics[width=\linewidth]{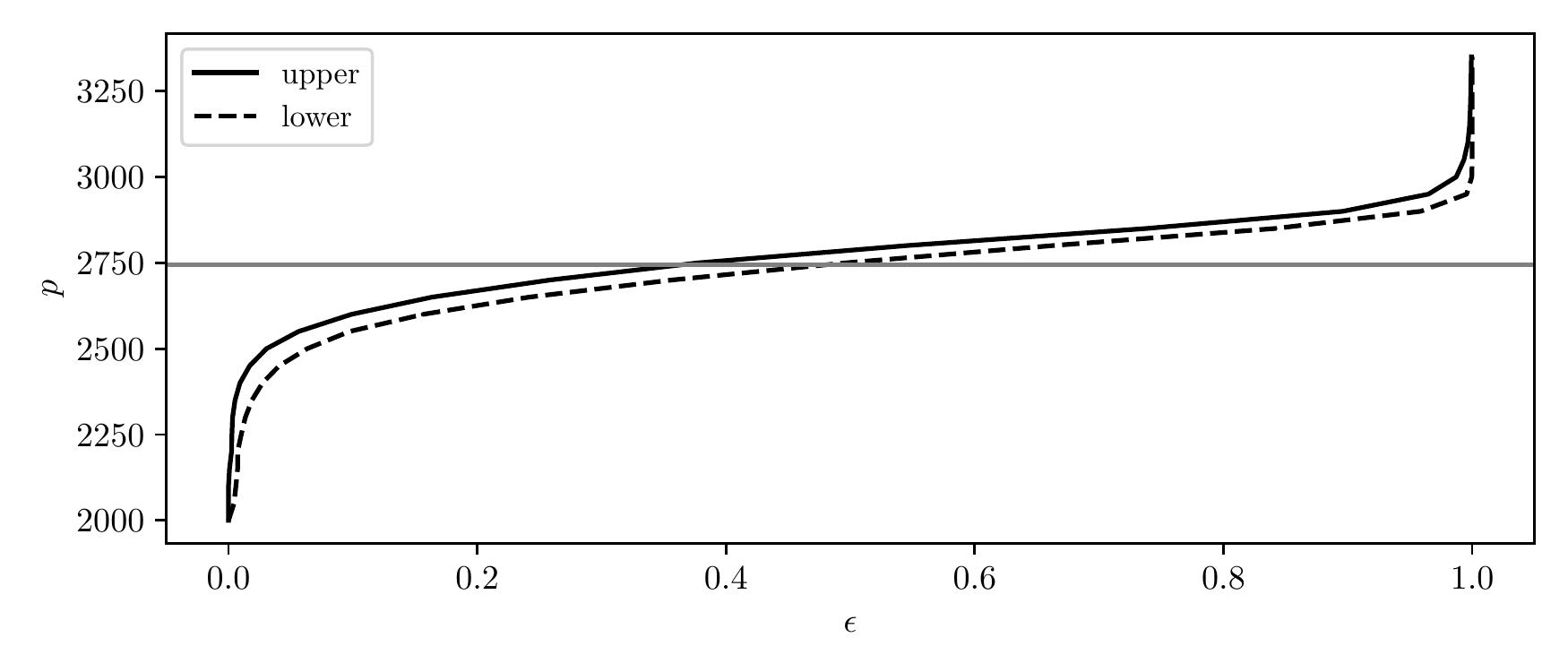}
  \caption{Bounds on $\mathbf{VaR}(p,\epsilon)$.}
  \label{fig:4}
\end{subfigure}\\

\medskip
\begin{subfigure}{0.8\textwidth}
  \includegraphics[width=\linewidth]{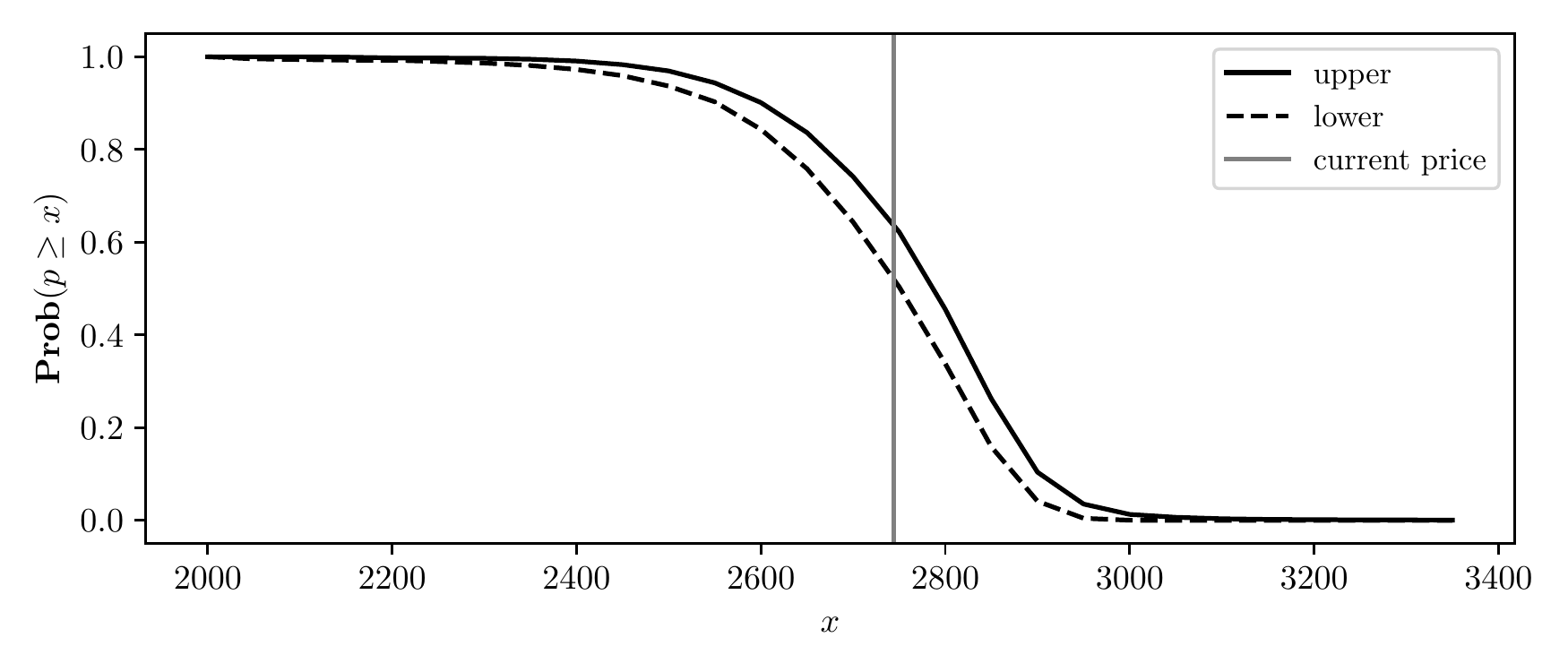}
  \caption{Bounds on $\Prob(p \geq x)$.}
  \label{fig:4}
\end{subfigure}

\caption{SPX example. Bounds on CDF, $\mathbf{VaR}$, and CCDF.}
\label{fig:sp_cdf_var}
\end{figure}

\begin{figure}
    \centering %
\begin{subfigure}{0.5\textwidth}
  \includegraphics[width=\linewidth]{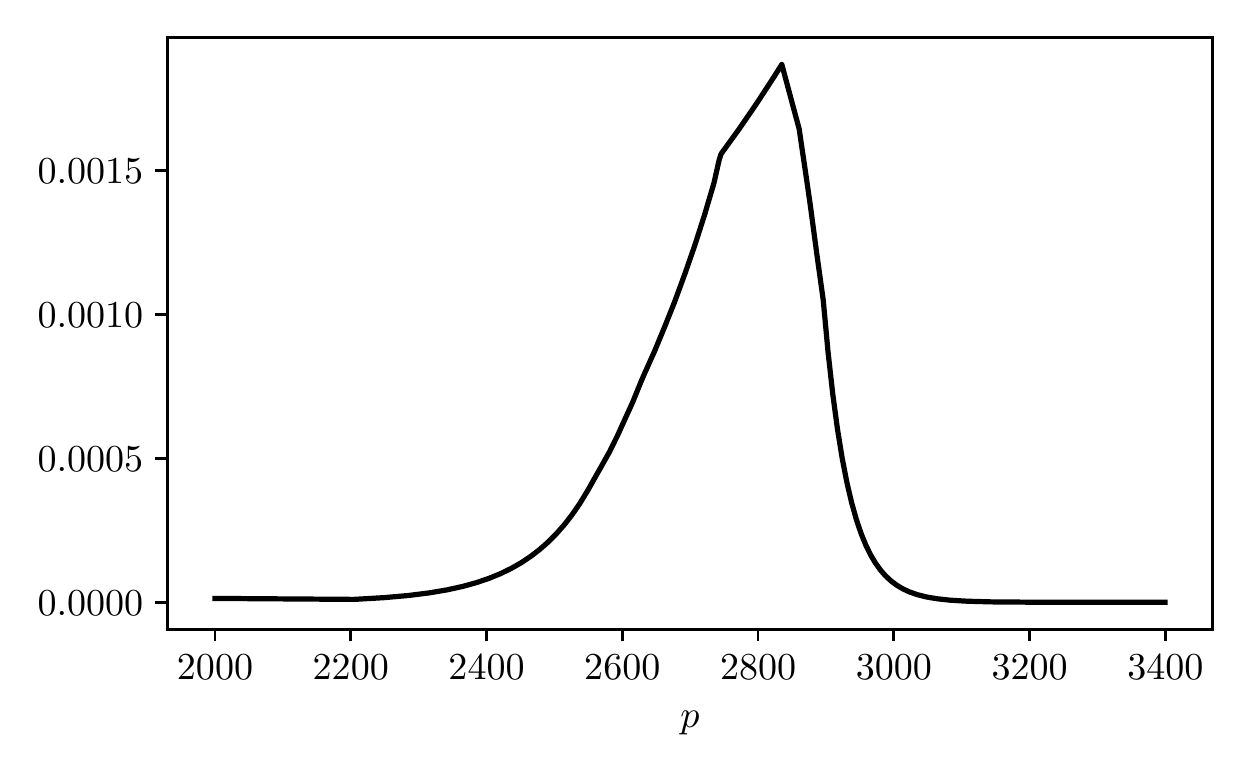}
\end{subfigure}\hfil %
\begin{subfigure}{0.5\textwidth}
  \includegraphics[width=\linewidth]{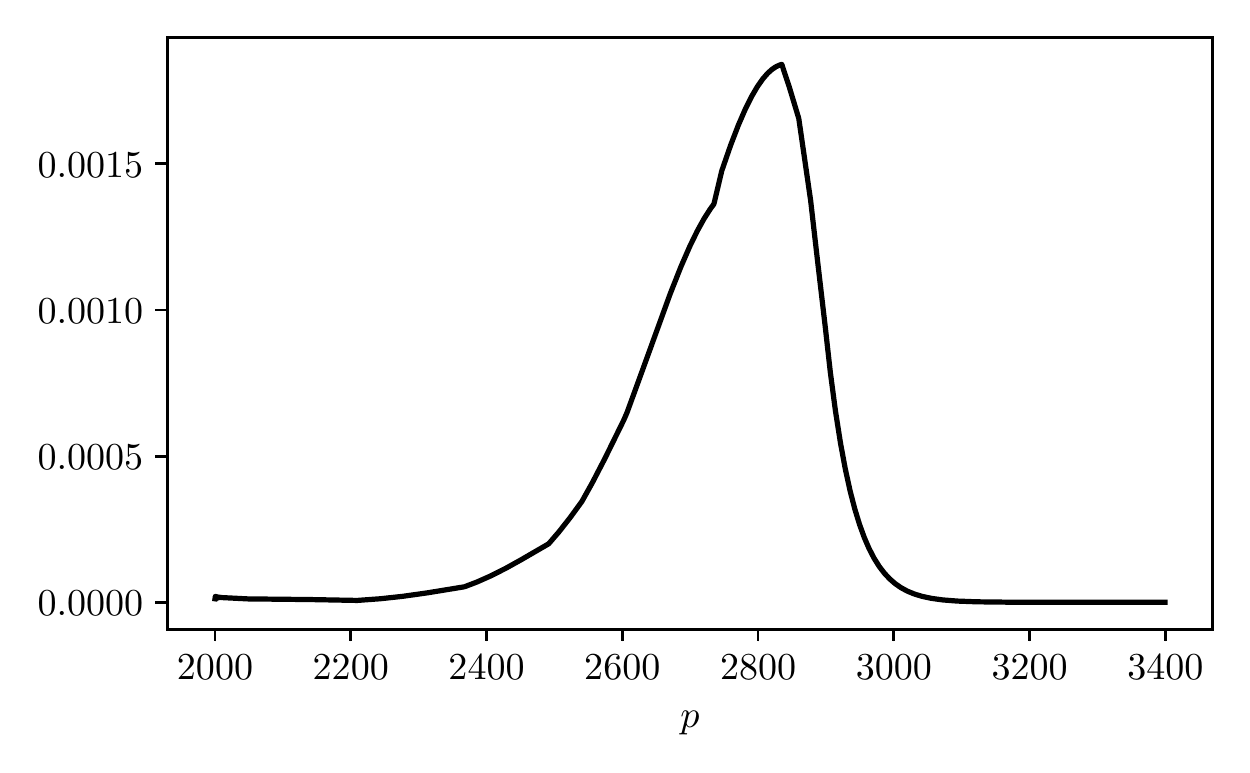}
\end{subfigure}
\caption{SPX example. Left: maximum entropy risk-neutral distribution; 
Right: closest log-normal distribution.}
\label{fig:sp_elements_risk_prob}
\end{figure}

\paragraph{Estimation.}
We computed the maximum entropy risk-neutral distribution,
as well as the (approximately) closest log-normal distribution to the set of risk-neutral probabilities.
The closest log-normal distribution was $\log(p) \sim \mathcal N(7.917, 0.05)$.
Via Monte-Carlo simulation, we found that the annualized volatility
of the index, assuming this log-normal distribution, was $19\%$,
which is on par with SPX's historical volatility of $15\%$.
The resulting distributions are visualized in figure~\ref{fig:sp_elements_risk_prob},
and appear to be heavy-tailed to the left, meaning there
is a large decrease in price is more probable than a large increase in price.

\begin{figure}
\begin{subfigure}{0.5\textwidth}
  \includegraphics[width=\linewidth]{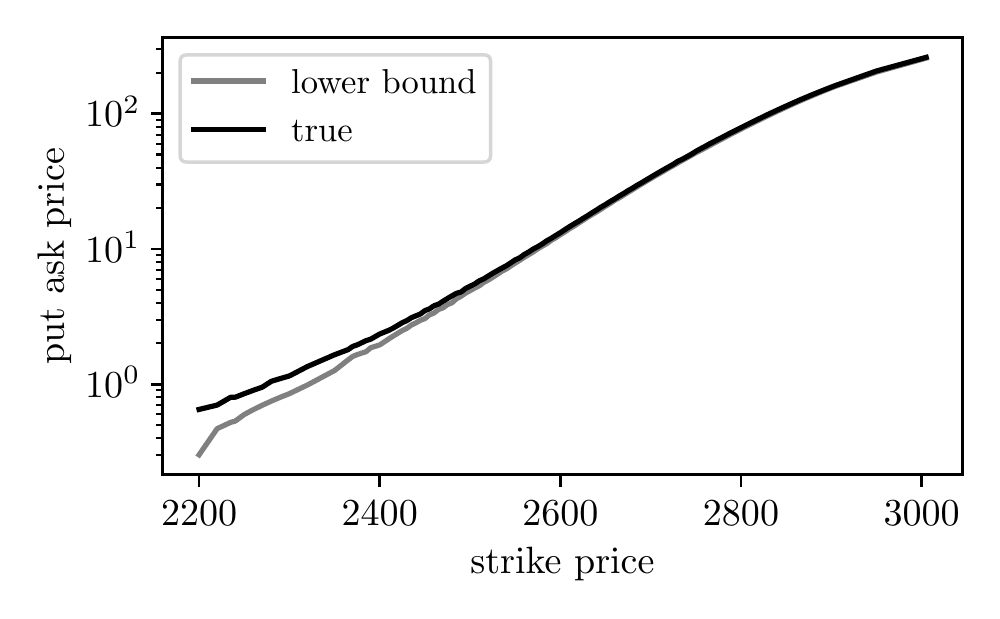}
  \caption{Put ask prices.}
  \label{fig:3}
\end{subfigure}\hfil %
\begin{subfigure}{0.5\textwidth}
  \includegraphics[width=\linewidth]{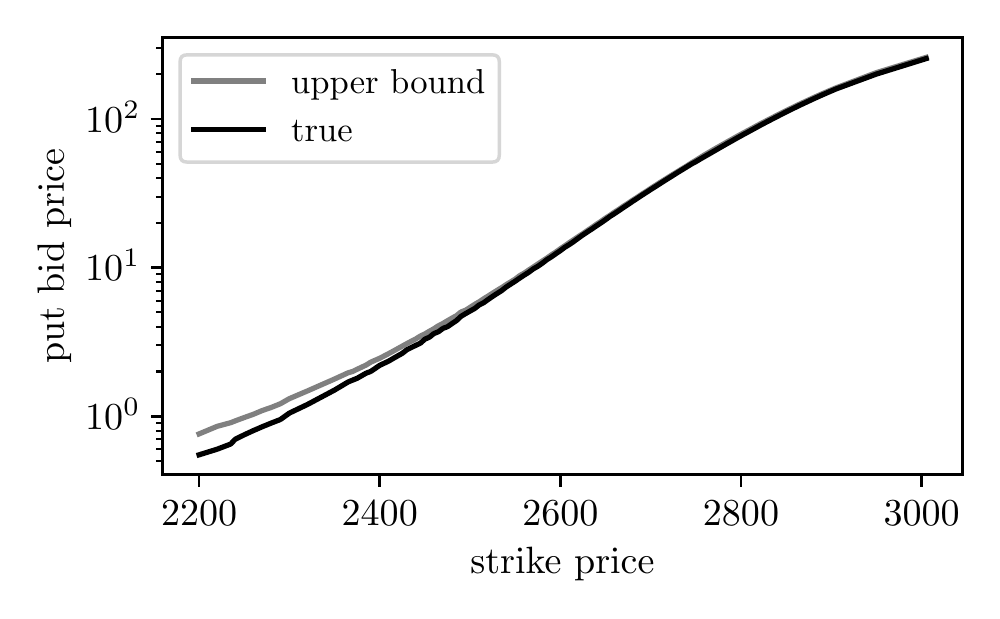}
  \caption{Put bid prices.}
  \label{fig:4}
\end{subfigure}\\

\medskip
\begin{subfigure}{0.5\textwidth}
  \includegraphics[width=\linewidth]{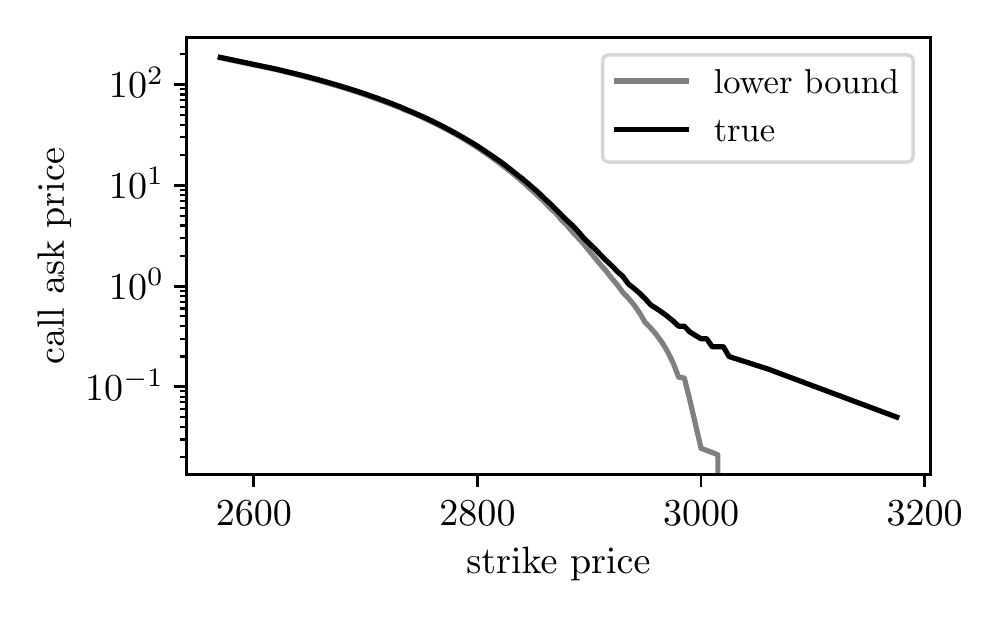}
  \caption{Call ask prices.}
  \label{fig:3}
\end{subfigure}\hfil %
\begin{subfigure}{0.5\textwidth}
  \includegraphics[width=\linewidth]{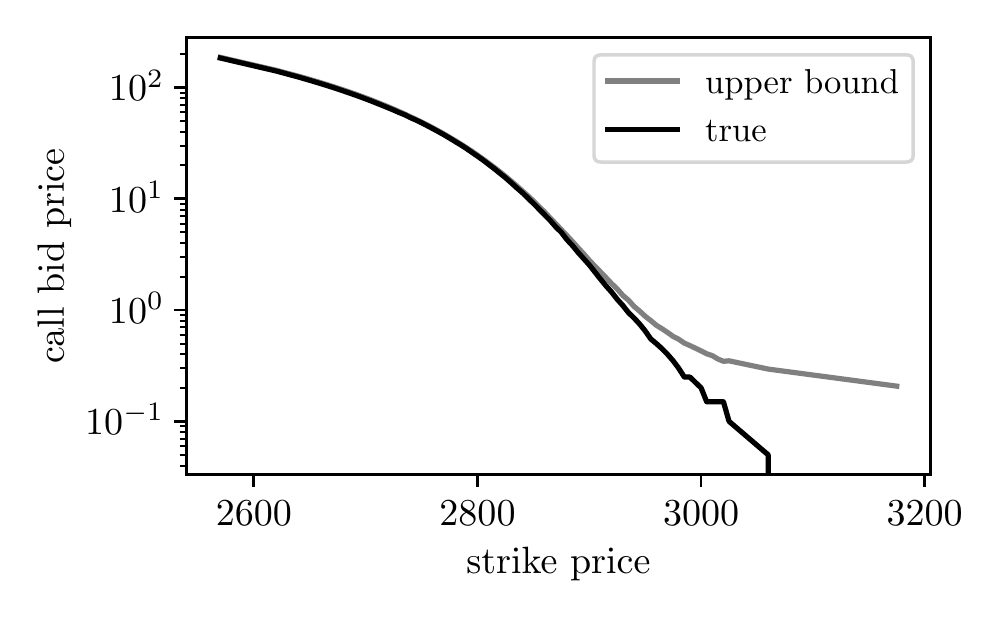}
  \caption{Call bid prices.}
  \label{fig:4}
\end{subfigure}
\caption{SPX example. Bounds on costs.}
\label{fig:sp_cost_bounds}
\end{figure}

\paragraph{Bounds on costs.}
We held out each put and call option one at a time
and computed bounds on their bid and ask prices.
In figure~\ref{fig:sp_cost_bounds} we plot our computed
lower and upper bounds along with the true prices.
We observe that the bounds seem to be quite tight,
and indeed bound the observed prices.

\subsection{Bitcoin}

In our next example, we consider
the crypto-currency Bitcoin as the underlying.
As derivatives, we use Deribit European-style options
and futures,
whose underlying is the Deribit BTC index,
which is the average of six leading
BTC-USD exchange prices: Bitstamp, Bittrex, Coinbase Pro, Gemini, Itbit,
and Kraken.
We gathered the prices of March 27, 2020 Bitcoin options
and futures on February 20, 2020 using the Deribit 
API \cite{deribit}.

We discretized the expiration price
from 5 to 30000 dollars, in 5 dollar increments,
resulting in $m=6000$ outcomes.
We allow six possible investments:
buying or selling puts, 
buying or selling calls, and
buying or selling futures.
The cost of each investment is
the ask price if buying, the negative bid price if selling,
plus a 0.04\% fee for option transactions,
a 0.075\% fee for buying futures,
and a 0.025\% (market-maker) rebate for selling futures
(which at the time of writing are the fees for the Deribit
exchange).

In total, there were 16 puts and 19 calls expiring
on March 27, with strike prices ranging from 4000 to 18000.
This means there were $n=2(16+19)+2=72$ possible investments.

\begin{figure}
\centering
\begin{subfigure}{0.8\textwidth}
  \includegraphics[width=\linewidth]{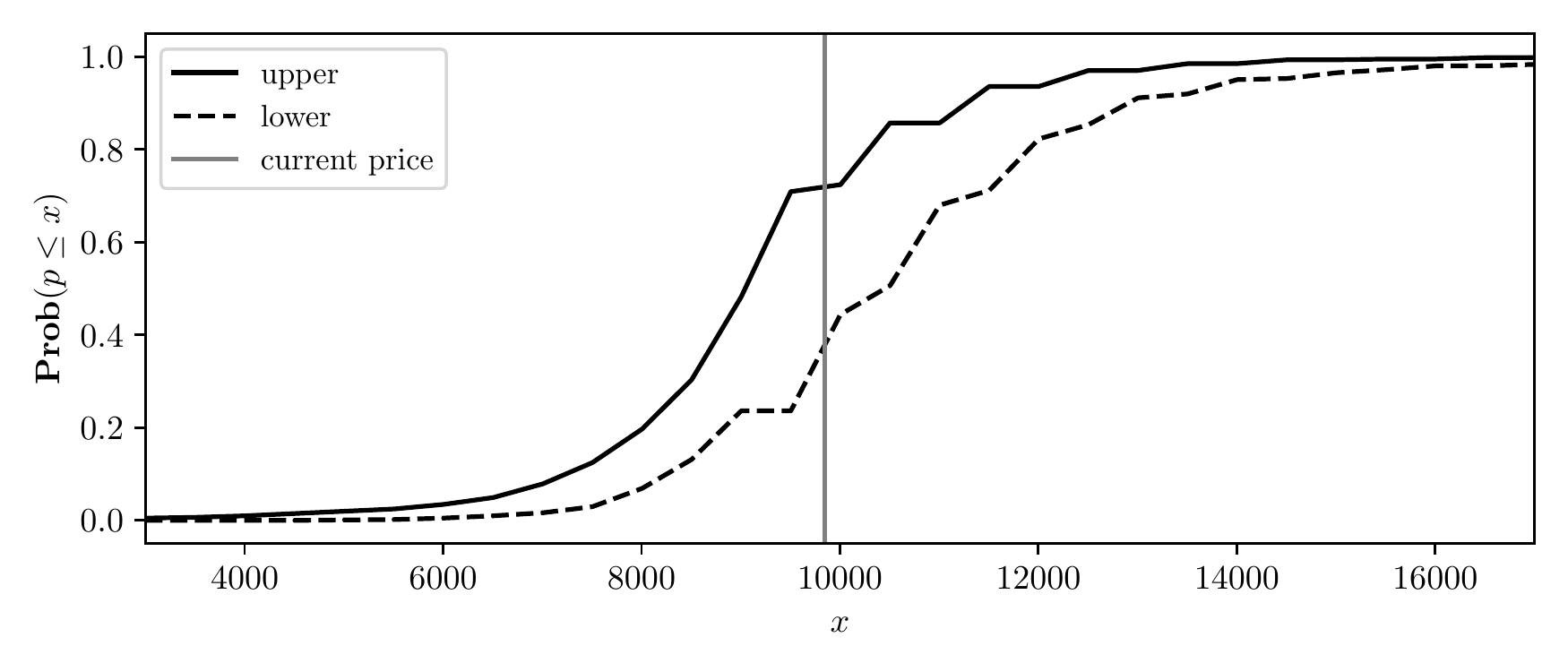}
  \caption{Bounds on $\Prob(p \leq x)$.}
  \label{fig:3}
\end{subfigure}\\

\medskip
\begin{subfigure}{0.8\textwidth}
  \includegraphics[width=\linewidth]{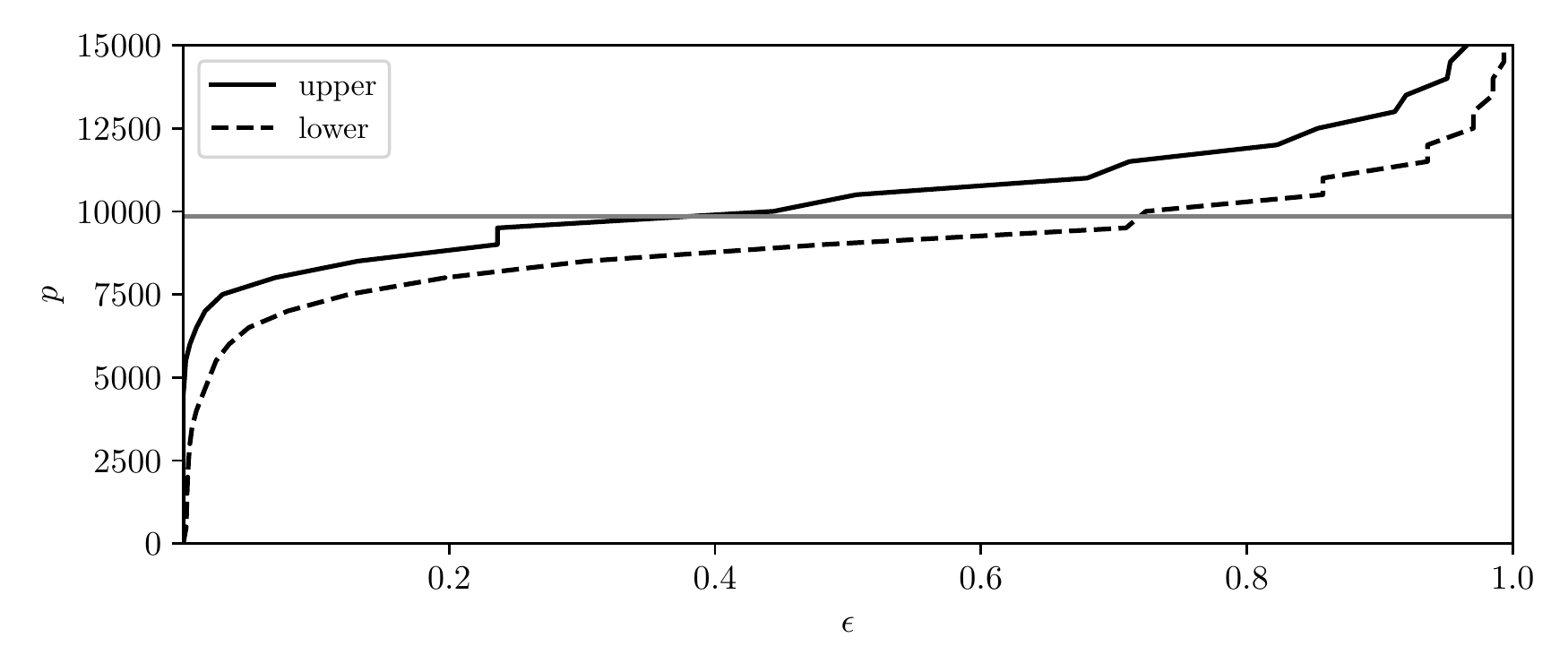}
  \caption{Bounds on $\mathbf{VaR}(p,\epsilon)$.}
  \label{fig:4}
\end{subfigure}\\

\medskip
\begin{subfigure}{0.8\textwidth}
  \includegraphics[width=\linewidth]{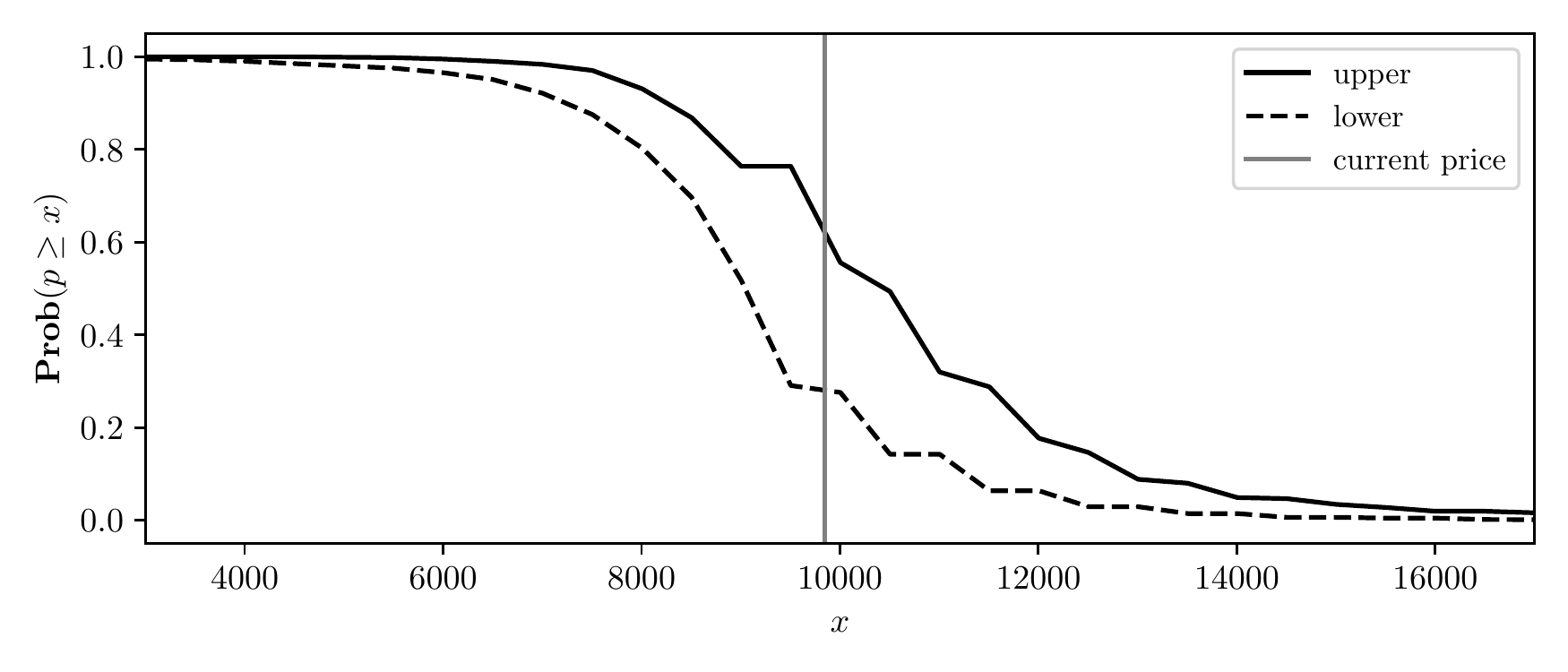}
  \caption{Bounds on $\Prob(p \geq x)$.}
  \label{fig:4}
\end{subfigure}

\caption{Bitcoin example. Bounds on CDF, $\mathbf{VaR}$, and CCDF.}
\label{fig:bitcoin_cdf_var}
\end{figure}

\paragraph{Functions of the price.}
We calculated bounds on the expected value of the expiration
price. The lower bound was 9847.7 dollars and the upper bound was 
9852.57 dollars.
We also computed bounds on the CDF, CCDF, and the value-at-risk 
of the expiration price.
In figure~\ref{fig:bitcoin_cdf_var} we plot these bounds.

\begin{figure}
    \centering %
\begin{subfigure}{0.5\textwidth}
  \includegraphics[width=\linewidth]{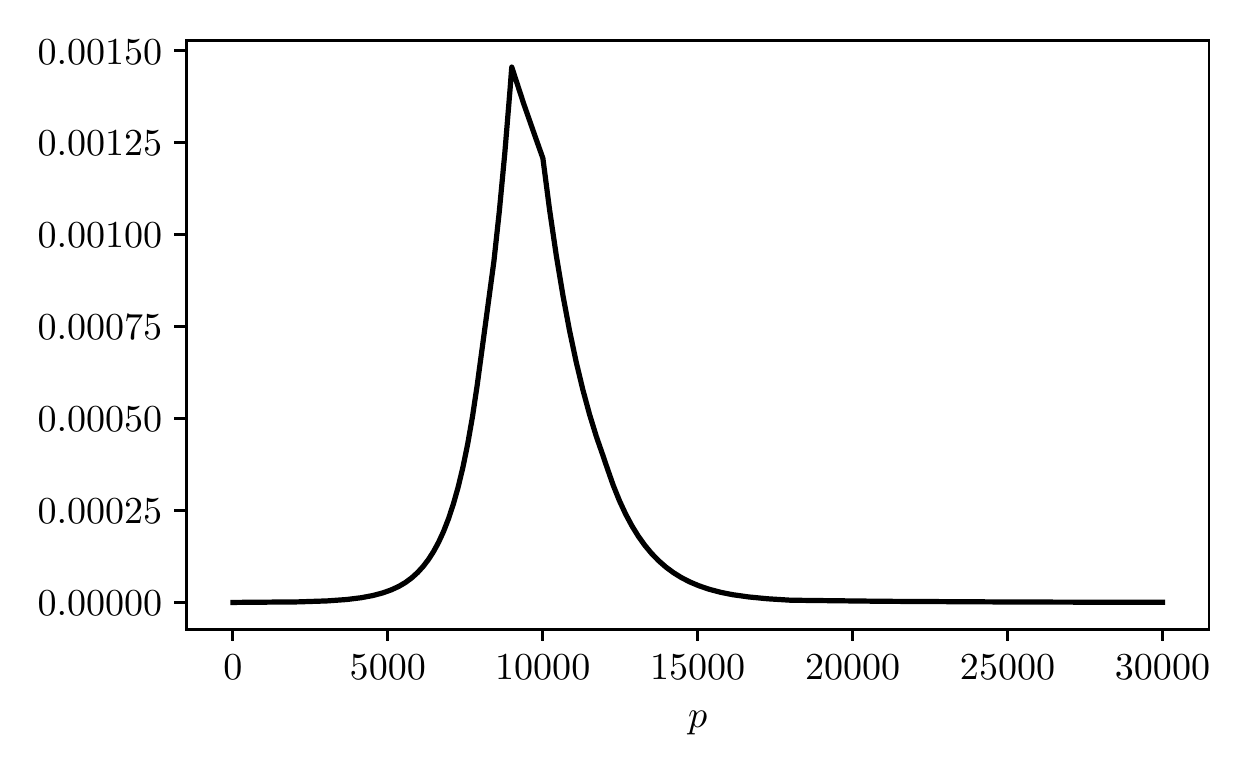}
\end{subfigure}\hfil %
\begin{subfigure}{0.5\textwidth}
  \includegraphics[width=\linewidth]{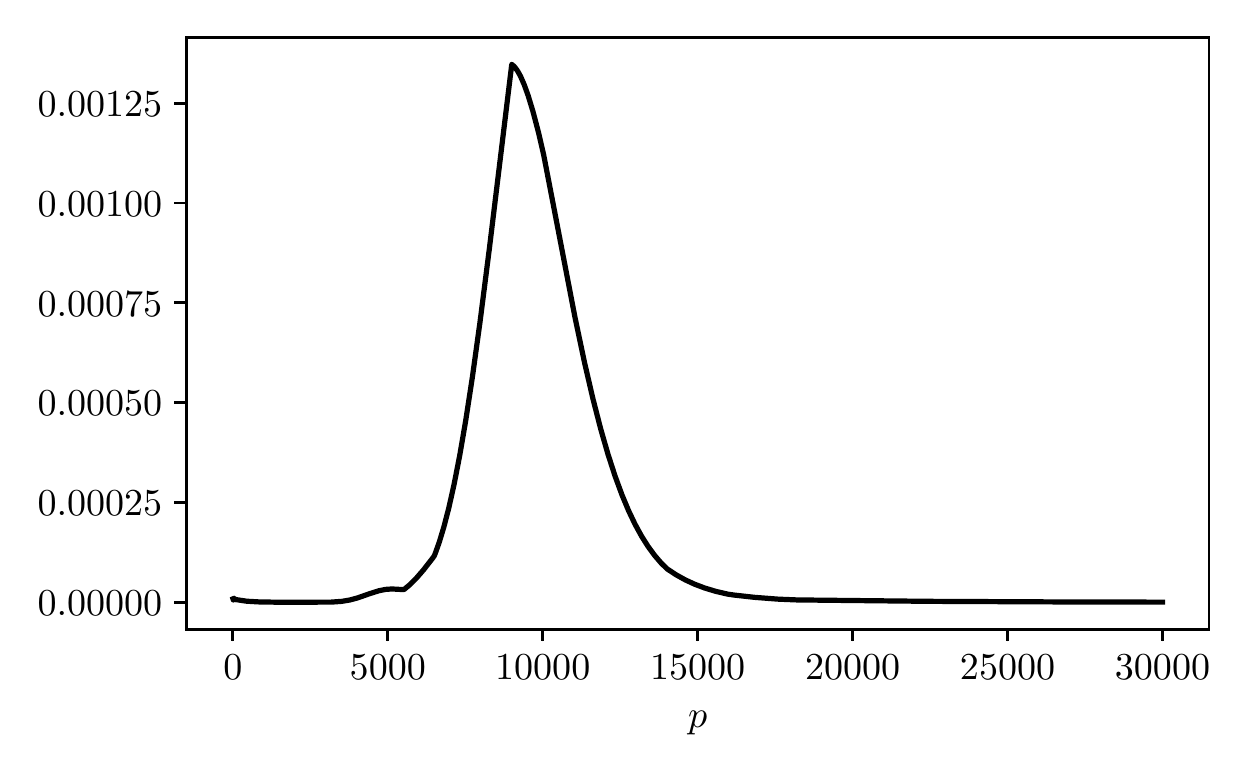}
\end{subfigure}
\caption{Bitcoin example. Left: maximum entropy risk-neutral distribution; Right: closest log-normal distribution.}
\label{fig:bitcoin_elements_risk_prob}
\end{figure}

\paragraph{Estimation.}
We computed the maximum entropy risk-neutral distribution,
as well as the (approximately) closest log-normal distribution to the set of risk-neutral probabilities.
The closest log-normal distribution was $\log(p) \sim \mathcal N(9.174, 0.204)$.
Via Monte-Carlo simulation, we found that the annualized volatility
of the index, assuming this log-normal distribution, was $71.8\%$.
The resulting distributions are visualized in figure~\ref{fig:bitcoin_elements_risk_prob},
and appear to be heavy-tailed to the right, which is the opposite
of the S\&P 500 example.

\begin{figure}
\begin{subfigure}{0.5\textwidth}
  \includegraphics[width=\linewidth]{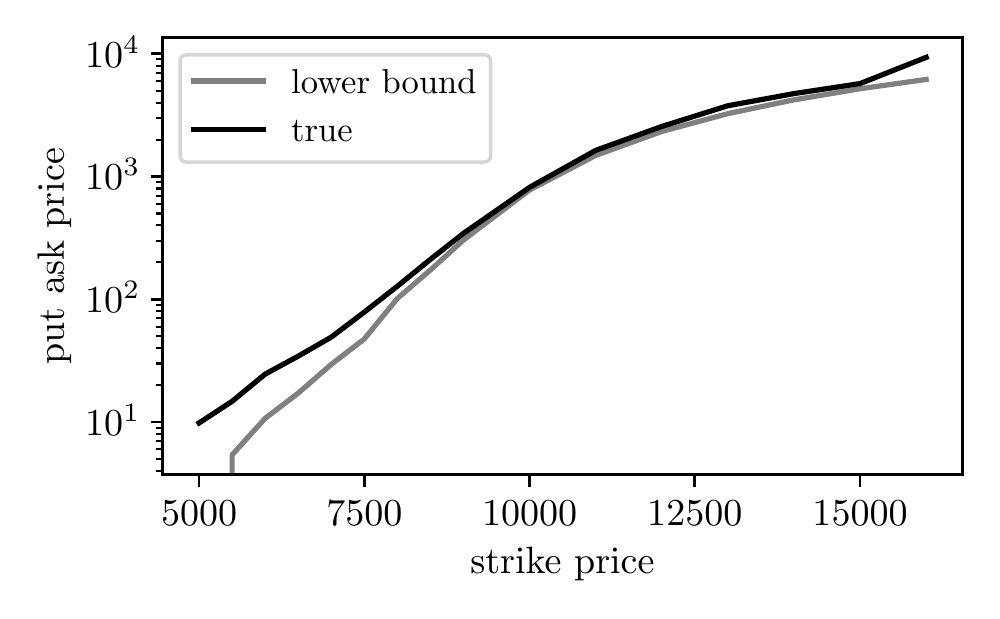}
  \caption{Put ask prices.}
  \label{fig:3}
\end{subfigure}\hfil %
\begin{subfigure}{0.5\textwidth}
  \includegraphics[width=\linewidth]{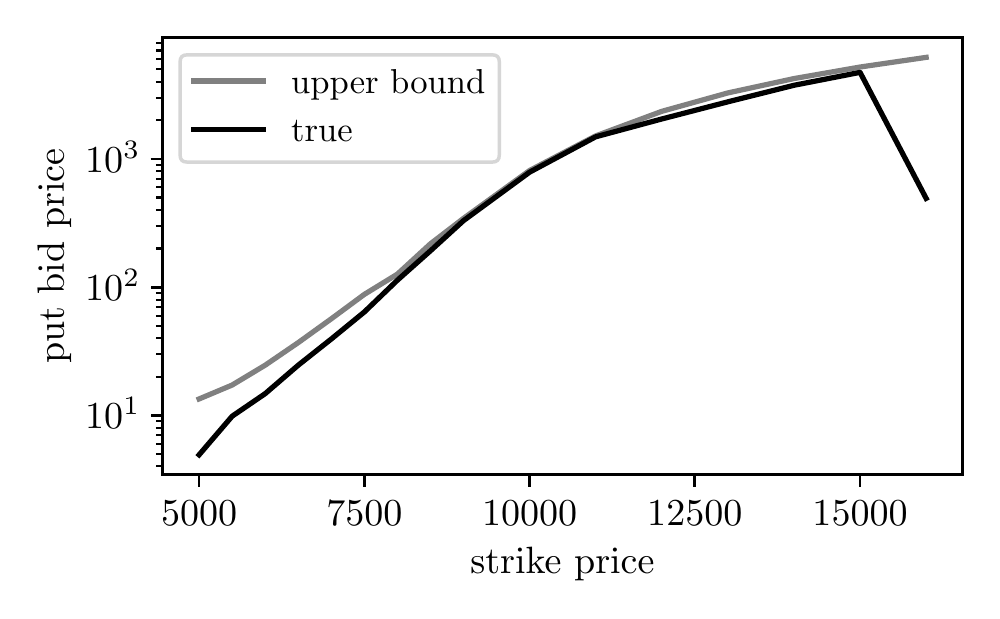}
  \caption{Put bid prices.}
  \label{fig:4}
\end{subfigure}\\

\medskip
\begin{subfigure}{0.5\textwidth}
  \includegraphics[width=\linewidth]{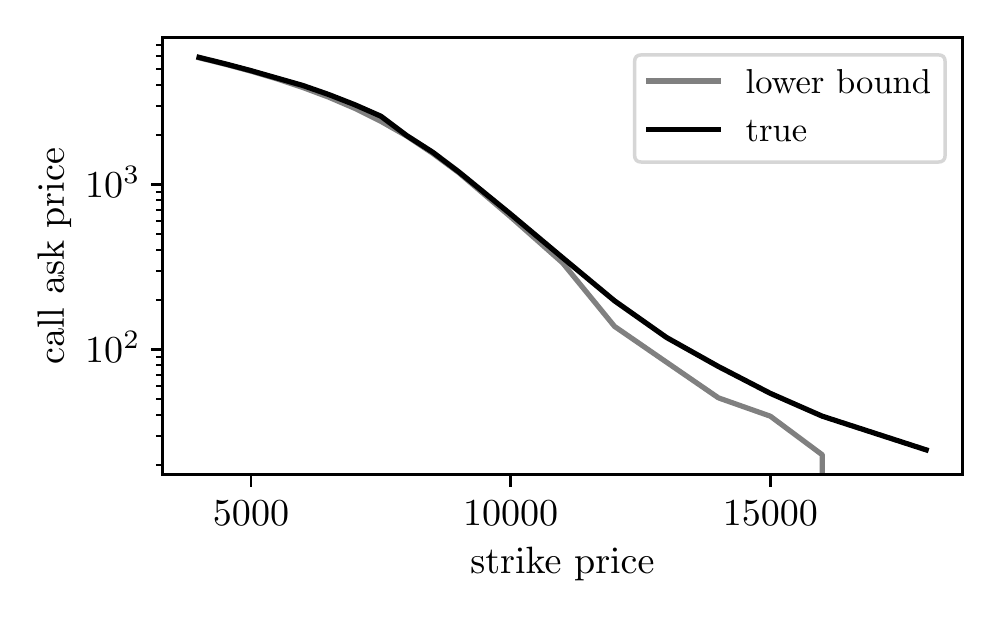}
  \caption{Call ask prices.}
  \label{fig:3}
\end{subfigure}\hfil %
\begin{subfigure}{0.5\textwidth}
  \includegraphics[width=\linewidth]{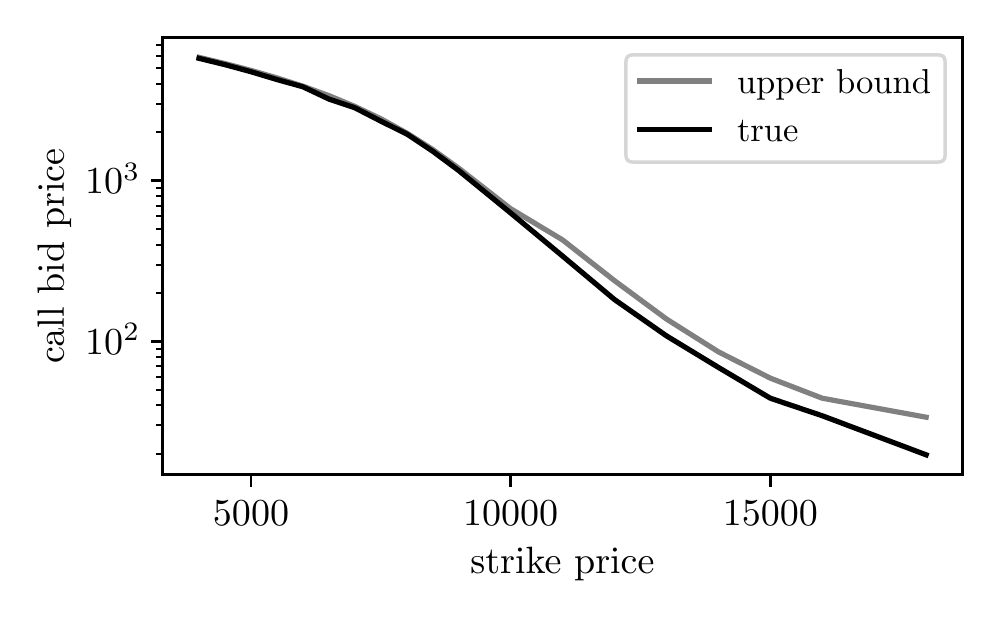}
  \caption{Call bid prices.}
  \label{fig:4}
\end{subfigure}
\caption{Bitcoin example. Bounds on costs.}
\label{fig:bitcoin_cost_bounds}
\end{figure}

\paragraph{Bounds on costs.}
We held out each put and call option one at a time
and computed bounds on their bid and ask prices.
In figure~\ref{fig:bitcoin_cost_bounds} we plot our computed
lower and upper bounds along with the true prices.
We observe that the bounds are quite tight,
and indeed bound the observed prices.

\paragraph{Sensitivities.}
We computed the optimal dual variable of the constraint $P^T\pi \leq c$
for the entropy maximization problem.
In table \ref{bitcoin_dual} we list the five largest dual variables,
along with their corresponding investments and costs.
We observe that shorting the underlying,
as well as buying/writing various calls
have the most effect on the maximum entropy risk-neutral probability.
For example, if we decrease the price of the 9000 call by ten dollars,
then the entropy will decrease by at least 0.01.

\begin{table}
  \caption{Bitcoin example. Dual variables for entropy maximization problem.}
  \label{bitcoin_dual}
  \centering
  \begin{tabular}{lll}
    \toprule
    Investment     & $c_i$     & $\lambda^\star_i$ \\
    \midrule
    Short underlying & -9847.65  & 0.001 \\
    Buy 9000 call     & 1191.268 & 0.001 \\
    Write 18000 call & -19.69       & 0.0004 \\
    Buy 10000 call     & 659.63       & 0.0004 \\
    Buy 8000 call & 1973.96       & 0.0002 \\
    \bottomrule
  \end{tabular}
\end{table}

\section{Conclusion}
In this paper we
described applications of minimizing
a convex or quasiconvex function over the set of
convex risk-neutral probabilities.
These include computation of bounds on the cumulative distribution,
VaR, conditional probabilities,
and prices of new derivatives, as well
as estimation problems.
We reiterate that all of the aforementioned
problems can be tractably solved,
and due to DSLs, are easy to implement.
A potential avenue for future research
is use the set
of risk-neutral probabilities
for multiple expiration dates
to somehow connect the distribution
of the underlying's price movements between those dates.

\clearpage
\section*{Acknowledgements}
Data from Wharton Research Data Services was used in preparing this paper.
S. Barratt is supported by the National Science Foundation Graduate Research Fellowship
under Grant No. DGE-1656518.
J. Tuck is supported by the Stanford Graduate Fellowship in Science \& Engineering.

\bibliography{refs}

\end{document}